\documentclass[sigplan,10pt]{acmart}

\usepackage[inline]{enumitem}
\usepackage{tikz}

\usetikzlibrary{fit,positioning,shapes.geometric,backgrounds,decorations.pathmorphing,decorations.markings,arrows.meta,shadows,decorations.pathreplacing}

\title{Netkit: Specializing Linux Packet Delivery for Container Networks}
\author{Daniel Borkmann}
\affiliation{
 \institution{Isovalent at Cisco}
 \country{Switzerland}
}
\email{daniel@iogearbox.net}
\author{Paul Chaignon}
\affiliation{
 \institution{Isovalent at Cisco}
 \country{France}
}
\email{paul.chaignon@gmail.com}
\authornote{The authors are in alphabetical order.}

\acmYear{2026}\copyrightyear{2026}
\setcopyright{cc}
\setcctype{by-nc-nd}
\acmConference[eBPF'26]{4th Workshop on eBPF and Kernel Extensions}{September 29-October 02, 2026}{Prague, Czech Republic}
\acmBooktitle{4th Workshop on eBPF and Kernel Extensions (eBPF'26), September 29-October 02, 2026, Prague, Czech Republic}
\acmDOI{10.1145/3837779.3838164}
\acmISBN{979-8-4007-2911-9/2026/09}



\begin{document}

\begin{abstract}
Cloud-native microservices architectures rely on network namespaces for isolation, with the overhead of container communications remaining a critical performance bottleneck.
While colocating containers on the same host mitigates some of this overhead, it cannot match the performance of communication within a single network namespace.
Existing solutions either require application rewrites or fail to support the full Linux network stack expected by containerized applications.

In this paper, we present \texttt{netkit}, an eBPF-based datapath that specializes the Linux networking stack to eliminate redundant backlog queue traversals during network namespace transitions.
netkit leverages eBPF to transparently redirect packets between namespaces, bypassing unnecessary buffering while preserving compatibility with existing container applications.
Our implementation in the Linux kernel, integrated with minimal changes to the Cilium network plugin for Kubernetes, improves throughput by up to 37\% and achieves parity between container-to-container and process-to-process communications, effectively closing the performance gap introduced by namespace isolation.
\end{abstract}

\settopmatter{printfolios=true}
\maketitle
\pagestyle{plain}

\section{Introduction}

Containerized deployments have become the de facto standard for cloud-native applications, with microservices architectures driving the distribution of application components across multiple containers.
As these distributed applications scale, the overhead of container-to-container communication becomes a critical performance bottleneck.
While colocating containers on the same host can mitigate some of this overhead, it is not always feasible, and even then it cannot match the performance of communication within a single network namespace.

The persistence of this overhead is particularly striking when considering the nature of network namespace switches.
Intuitively, transitions between network namespaces should not incur any performance penalty, as they represent logical rather than physical boundaries.
While copying data is necessary to enforce memory isolation when crossing the kernel-userspace boundary, no such requirement exists for network namespace boundaries.

Significant work has gone into studying and improving the performance of the Linux networking stack, with recent efforts focusing on container networks as they become more prevalent. eBPF has emerged as a key technology in this space, offering new ways to hook into the stack \cite{xdp, ovs-afxdp, netedit} and enabling the replacement of legacy algorithms with more efficient implementations \cite{katran, bpfilter, hermes}. In the context of container networks, eBPF has been used to replace and specialize almost all aspects of the stack, from load balancing and policy enforcement \cite{cilium} to quality of service (QoS) mechanisms \cite{beeqos} and proxy redirection \cite{hybridmesh, spright}.

However, even with BPF's capabilities, the underlying packet delivery mechanism in Linux remains generic and unoptimized for the specific use case of container-to-container communication.
As we show in Section \ref{sec:background-evals}, for short-lived connections, two processes in the same network namespace\footnote{We use two processes in the host network namespace, with communication over the loopback device.} achieve 31\% higher throughput than two containers on the same host.
Likewise, two hosts achieve 26\% higher throughput than two containers communicating over the wire.
This gap leaves a critical performance improvement untapped.

Several new packet delivery mechanisms based on BPF have been proposed in the past.
AF\_XDP implements a form of partial kernel bypass, where an XDP program can send packets directly to userspace \cite{ovs-afxdp}.
However, container applications typically listen on IPv4/IPv6 sockets, so AF\_XDP either requires traffic to be reinjected into the kernel or the application to be rewritten.
sockmap implements a socket-level redirection to exchange messages between IPv4/IPv6 sockets using a special BPF socket map \cite{spright, hybridmesh}, but it only covers a subset of use cases, as applications often need to send and receive traffic from physical devices on the host.
Both of these mechanisms are ultimately a poor fit for applications that expect a full Linux network stack in their network namespace, especially as such applications are becoming more common with containerized virtual machines, such as Kata Containers and KubeVirt.

In this paper, we present \texttt{netkit}, a high-performance eBPF-based datapath for containers.
netkit specializes the packet delivery mechanisms in the Linux network stack to eliminate all overhead associated with network namespace transitions.
It leverages eBPF to implement routing and any required packet processing logic in the host network namespace, such as access control and traffic shaping.
Crucially, netkit operates transparently to containerized applications and requires minimal modifications to integrate with existing BPF-based container datapaths, making it a practical solution for real-world deployments.

We have contributed netkit to the Linux kernel \cite{redirectpeer-patch, netkit-patch} and demonstrate that, with limited changes to the Cilium Kubernetes CNI, it improves throughput by up to 37\% and matches process-to-process communications in both throughput and latency.

\section{Background}

\label{sec:background}

The Linux networking stack may employ a per-CPU backlog queue \cite{backlog-queue} to buffer incoming packets before they are processed by the protocol stack.
Each CPU core maintains its own queue, allowing packets received by network devices to be distributed across cores, improving scalability on multicore systems.
When a backlog queue is used, network device drivers enqueue received packets to the backlog queue of the target CPU, where they are later processed by the upper networking stack.

\begin{figure}
  \centering
  \begin{tikzpicture}[
    layer/.style={draw, thick, fill=white, drop shadow, minimum width=2.5cm, minimum height=0.5cm, align=center},
    backlog/.style={draw, thick, decorate, decoration={zigzag,segment length=3pt}},
    sidelabel/.style={rectangle, text width=2cm, text centered, minimum height=0.5cm, rotate=-90}
  ] 
    \node[layer, label={[font=\small, label distance=0.3cm]right:sockmap}] (socket) {Socket};
    \node[layer, anchor=north] at (socket.south) (tcp) {TCP};
    \node[layer, anchor=north] at (tcp.south) (netfilter) {Netfilter \& IP};
    \node[layer, anchor=north, label={[font=\small, label distance=0.3cm]right:tc-bpf}] at (netfilter.south) (tc) {Traffic Control};
    \node[layer, anchor=north, label={[font=\small, label distance=0.3cm]right:XDP}] at ([yshift=-0.8cm]tc.south) (driver) {Driver};

    \draw[decoration={brace,amplitude=5pt,raise=5pt},decorate, thick] (tc.south west) -- (socket.north west);
    \node[sidelabel, anchor=north] at ([xshift=-0.5cm]tcp.south west) {Upper stack};

    \draw[backlog] (driver.north) -- ([yshift=0.4cm]driver.north);
    \draw[thick, -Triangle] ([yshift=0.4cm]driver.north) -- (tc.south);

    \draw[thick] (socket.east) -- ([xshift=0.3cm]socket.east);
    \draw[thick] (tc.east) -- ([xshift=0.3cm]tc.east);
    \draw[thick] (driver.east) -- ([xshift=0.3cm]driver.east);
  \end{tikzpicture}
  \Description{Illustration of the usual TCP/IP networking stack and the main BPF networking hooks in Linux, on the receive side when traversing a backlog queue (illustrated with zigzags).}
  \caption{Illustration of the usual TCP/IP networking stack and the main BPF networking hooks in Linux, on the receive side when traversing a backlog queue (illustrated with zigzags).}
  \label{fig:linux-networking-stack}
\end{figure}
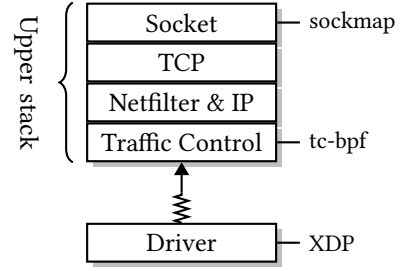

Figure \ref{fig:linux-networking-stack} illustrates the boundary between the driver and the upper stack.
The majority of packet processing occurs in the upper stack, as it offers greater flexibility to implement a wide range of network features.
For this same reason, container networking solutions typically hook into the upper stack, either using netfilter or the tc-bpf hook, to introduce custom packet handling, filtering, or forwarding logic.

Packets received from the wire on a physical device typically do not go through the backlog queue\footnote{Unless Receive Packet Steering (RPS), the software counterpart of RSS, is enabled.}.
However, Virtual Ethernet (veth) devices, used to connect container network namespaces to the host, utilize the per-CPU backlog queue for packet processing.
veth devices operate as interconnected pairs, with one device residing in the container's network namespace and the other in the host namespace.
When a packet is transmitted from the container, it is received by the host device and enqueued to the per-CPU backlog queue of the target CPU for processing, mirroring the behavior of physical network devices.
The same occurs in reverse via the veth pair.

\def\pathid#1{\tikz[baseline]{\node[draw, thick, circle, anchor=base, font=\footnotesize] {#1};}}

\begin{figure}
  \centering
  \begin{tikzpicture}[
    netns/.style={draw, thick, rounded corners, fill=white, drop shadow, minimum width=2cm, minimum height=1.5cm, align=center},
    hostns/.style={draw, thick, rounded corners, fill=white, drop shadow, minimum width=5.5cm, minimum height=4.5cm, align=center},
    device/.style={draw, thick, fill=white, drop shadow, minimum width=0.4cm, minimum height=0.4cm},
    pathid/.style={draw, thick, circle, fill=white, drop shadow, font=\footnotesize},
    pktpath/.style={thick, rounded corners=7pt},
    redirect/.style={thick, rounded corners=7pt, densely dashed, -Triangle},
    backlog/.style={thick, decorate, decoration={zigzag,segment length=3pt}}
  ]
    \node[hostns] (host1) {};
    \node[netns, anchor=north west, label={[font=\small \bfseries]above:Container 1}] at ([xshift=0.5cm,yshift=-0.5cm]host1.north west) (container1) {};
    \node[netns, anchor=north east, label={[font=\small \bfseries]above:Container 2}] at ([xshift=-0.5cm,yshift=-0.5cm]host1.north east) (container2) {};
    \node[device, anchor=south] at (container1.south) (veth1) {};
    \node[device, anchor=north, label={[font=\scriptsize \bfseries]right:veth pair}] at (container1.south) (lxc1) {};
    \node[device, anchor=south] at (container2.south) (veth2) {};
    \node[device, anchor=north, label={[font=\scriptsize \bfseries]right:veth pair}] at (container2.south) (lxc2) {};
    \node[device, anchor=south, label={[font=\scriptsize \bfseries]right:Physical device}] at (host1.south) (eth1) {};

    \node[hostns, anchor=west, minimum width=2.6cm] at ([xshift=0.3cm]host1.east) (host2) {};
    \node[netns, anchor=north, label={[font=\small \bfseries]above:Container 3}] at ([yshift=-0.5cm]host2.north) (container3) {};
    \node[device, anchor=south] at ([xshift=-0.2cm]container3.south) (veth3) {};
    \node[device, anchor=north, label={[font=\scriptsize \bfseries]right:veth pair}] at ([xshift=-0.2cm]container3.south) (lxc3) {};
    \node[device, anchor=south] at ([xshift=-0.2cm]host2.south) (eth2) {};

    \draw[pktpath] ([yshift=0.4cm]veth1.north) -- (lxc1.south);
    \draw[backlog] (lxc1.south) -- ([yshift=-0.4cm]lxc1.south);
    \begin{scope}[decoration={markings, mark=between positions 0.2 and 0.8 step 0.35 with {\arrow{Triangle}}}]
      \draw[pktpath, rounded corners=10pt, postaction={decorate}] ([yshift=-0.4cm]lxc1.south) -- ([yshift=-0.8cm]lxc1.south) -- ([yshift=-0.8cm]lxc2.south) -- (veth2.north);
    \end{scope}
    \draw[backlog] (veth2.north) -- ([yshift=0.4cm]veth2.north);
    \draw[pktpath, -Triangle] ([yshift=0.4cm]veth2.north) -- ([yshift=0.7cm]veth2.north);

    \draw[pktpath] ([yshift=0.4cm]veth3.north) -- (lxc3.south);
    \draw[backlog] (lxc3.south) -- ([yshift=-0.4cm]lxc3.south);
    \begin{scope}[decoration={markings, mark=between positions 0.15 and 1 step 0.35 with {\arrow{Triangle}}}]
      \draw[pktpath, rounded corners=10pt, postaction={decorate}] ([yshift=-0.4cm]lxc3.south) |- ([yshift=-0.5cm]eth1.south) -- (eth1.north);
    \end{scope}
    \draw[pktpath] (eth1.north) -- ([yshift=0.4cm]eth1.north);
    \draw[pktpath, rounded corners=10pt] ([yshift=0.4cm]eth1.north) |- ([xshift=-1cm,yshift=-0.8cm]lxc2.south);

    \node[pathid] at ([xshift=0.35cm,yshift=0.35cm]eth1.north east) {1};
    \node[pathid] at ([xshift=-0.3cm,yshift=-0.9cm]lxc1.south) {2};
    \node[pathid] at ([xshift=0.3cm,yshift=0.7cm]eth2.north east) {3};

    \node[anchor=west, align=left] at ([xshift=0.7cm,yshift=-1cm]host1.south west) (legend) {Packet path\\Backlog queue traversal};
    \draw[pktpath, -Triangle] ([xshift=-0.5cm,yshift=0.25cm]legend.west) -- ([yshift=0.25cm]legend.west);
    \draw[backlog] ([xshift=-0.5cm,yshift=-0.25cm]legend.west) -- ([yshift=-0.25cm]legend.west);
  \end{tikzpicture}
  \Description{Container-to-container packet paths through the Linux stack and veth device pairs, on a single host and over the wire.}
  \caption[Container-to-container packet paths through the Linux stack and veth device pairs, on a single host and over the wire.]{Container-to-container packet paths through the Linux stack and veth device pairs, on a single host (\pathid{2}) and over the wire (\pathid{3} and \pathid{1}).}
  \label{fig:vanilla-packet-paths}
\end{figure}
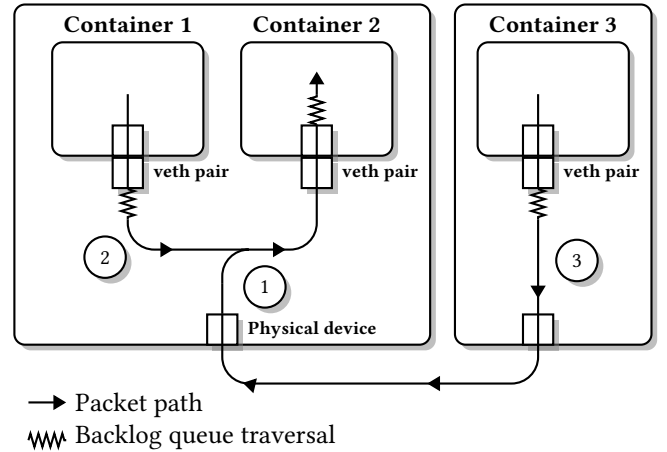

As a result, as shown in Figure \ref{fig:vanilla-packet-paths}, a packet traveling from one container to another on the same host (path \pathid{2}) will traverse the per-CPU backlog queue twice.
Similarly, a packet traveling from the wire to a container (path \pathid{1}) (or vice versa, path \pathid{3}) will also pass through the backlog queue at the destination container.

In addition, in the veth case, processing of packets from the backlog queue may be deferred to a dedicated kernel thread, ksoftirqd, when under load.
This thread runs at normal scheduling priority and competes with all other runnable tasks on the CPU, which can significantly increase latency.
It also distorts scheduler accounting, as the cycles spent delivering the packet are no longer charged to the process that sent it.

\label{sec:background-evals}

To measure this overhead, we run the netperf \texttt{TCP\_CRR} benchmark between containerized processes across four distinct configurations:
\begin{enumerate*}[label = (\arabic*)]
  \item same host, no network namespaces,
  \item same host, with network namespaces,
  \item different hosts, no network namespaces, and
  \item different hosts, with network namespaces.
\end{enumerate*}
In the first scenario, with the client and server processes in the same network namespace on the same host, communication occurs over the loopback device.

For all other configurations, the processes use veth devices to establish connectivity.
All tests run on an Intel Xeon CPU clocked at 3.1GHz, with 64KB, 2MB, and 24.8MB L1, L2, and L3 CPU caches.
The tests are not limited by either the available memory or the Ethernet adapters' capacity.

\begin{figure}
  \centering
  \includegraphics[width=\linewidth]{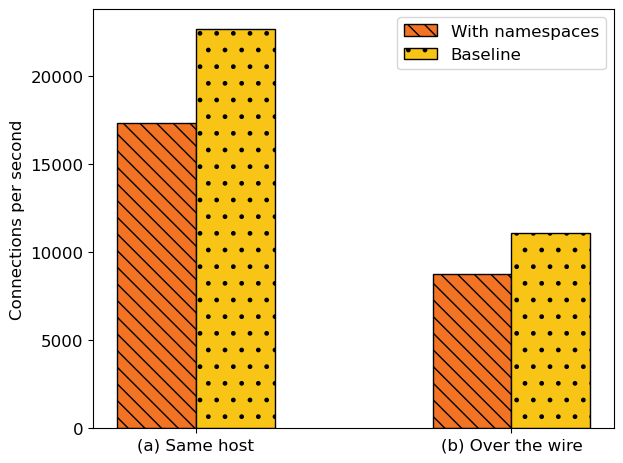}
  \Description{Overhead of network namespace transitions for containers on the same host and on different hosts, across the wire.}
  \caption{Overhead of network namespace transitions for containers on the same host and on different hosts, across the wire.}
  \label{fig:benchmark}
\end{figure}

The results, presented in Figure \ref{fig:benchmark}, reveal a notable throughput penalty when processes communicate across network namespace boundaries.
Specifically, removing the network namespaces increases throughput by 31\% for two processes on the same host and by 26\% for a connection over the wire.

\section{Design}

\begin{figure}
  \centering
  \begin{tikzpicture}[
    netns/.style={draw, thick, rounded corners, fill=white, drop shadow, minimum width=2cm, minimum height=1.5cm, align=center},
    hostns/.style={draw, thick, rounded corners, fill=white, drop shadow, minimum width=5.5cm, minimum height=4.5cm, align=center},
    device/.style={draw, thick, fill=white, drop shadow, minimum width=0.4cm, minimum height=0.4cm},
    pathid/.style={draw, thick, circle, fill=white, drop shadow, font=\footnotesize},
    pktpath/.style={thick, rounded corners=7pt},
    redirect/.style={thick, rounded corners=7pt, densely dashed, -Triangle},
    backlog/.style={thick, decorate, decoration={zigzag,segment length=3pt}}
  ]
    \node[hostns] (host1) {};
    \node[netns, anchor=north west, label={[font=\small \bfseries]above:Container 1}] at ([xshift=0.5cm,yshift=-0.5cm]host1.north west) (container1) {};
    \node[netns, anchor=north east, label={[font=\small \bfseries]above:Container 2}] at ([xshift=-0.5cm,yshift=-0.5cm]host1.north east) (container2) {};
    \node[device, anchor=south] at ([xshift=0.1cm]container1.south) (veth1) {};
    \node[device, anchor=north, label={[font=\scriptsize \bfseries]left:netkit pair}] at ([xshift=0.1cm]container1.south) (lxc1) {};
    \node[device, anchor=south] at ([xshift=-0.1cm]container2.south) (veth2) {};
    \node[device, anchor=north, label={[font=\scriptsize \bfseries]right:netkit pair}] at ([xshift=-0.1cm]container2.south) (lxc2) {};
    \node[device, anchor=south, label={[font=\scriptsize \bfseries]right:Physical device}] at (host1.south) (eth1) {};

    \node[hostns, anchor=west, minimum width=2.6cm] at ([xshift=0.3cm]host1.east) (host2) {};
    \node[netns, anchor=north, label={[font=\small \bfseries]above:Container 3}] at ([yshift=-0.5cm]host2.north) (container3) {};
    \node[device, anchor=south] at ([xshift=-0.3cm]container3.south) (veth3) {};
    \node[device, anchor=north, label={[font=\scriptsize \bfseries]right:netkit pair}] at ([xshift=-0.3cm]container3.south) (lxc3) {};
    \node[device, anchor=south] at ([xshift=-0.3cm]host2.south) (eth2) {};

    \draw[pktpath] ([yshift=0.4cm]veth1.north) -- (lxc1.north);
    \draw[redirect] (lxc1.north) -- ([yshift=-0.1cm]lxc1.north) -- ([yshift=-0.4cm]lxc2.south) -- (lxc2.south);
    \draw[pktpath] (lxc2.south) -- (veth2.north);
    \draw[backlog] (veth2.north) -- ([yshift=0.4cm]veth2.north);
    \draw[pktpath, -Triangle] ([yshift=0.4cm]veth2.north) -- ([yshift=0.7cm]veth2.north);

    \draw[pktpath] ([yshift=0.4cm]veth3.north) -- (lxc3.north);
    \draw[redirect] (lxc3.north) -- (eth2.north);
    \begin{scope}[decoration={markings, mark=between positions 0.3 and 1 step 0.35 with {\arrow{Triangle}}}]
      \draw[pktpath, rounded corners=10pt, postaction={decorate}] (eth2.north) |- ([yshift=-0.5cm]eth1.south) -- (eth1.north);
    \end{scope}
    \draw[pktpath] (eth1.north) -- ([yshift=0.4cm]eth1.north);
    \draw[redirect] ([yshift=0.4cm]eth1.north) -- ([yshift=0.8cm]eth1.north) -- (veth2.center) -- ([xshift=0.5cm,yshift=0.7cm]veth2.north);

    \node[pathid] at ([xshift=0.35cm,yshift=0.5cm]eth1.north east) {1};
    \node[pathid] at ([xshift=0.45cm,yshift=-0.35cm]lxc1.south) {2};
    \node[pathid] at ([xshift=0.3cm,yshift=0.7cm]eth2.north east) {3};

    \node[anchor=west, align=left] at ([xshift=0.7cm,yshift=-1cm]host1.south west) (legend) {Packet path\\Backlog queue traversal\\BPF redirect};
    \draw[pktpath, -Triangle] ([xshift=-0.5cm,yshift=0.4cm]legend.west) -- ([yshift=0.4cm]legend.west);
    \draw[backlog] ([xshift=-0.5cm]legend.west) -- (legend.west);
    \draw[redirect] ([xshift=-0.5cm,yshift=-0.4cm]legend.west) -- ([yshift=-0.4cm]legend.west);
  \end{tikzpicture}
  \Description{Container-to-container packet paths using BPF redirects and netkit device pairs, on a single host and over the wire.}
  \caption[Container-to-container packet paths using BPF redirects and netkit device pairs, on a single host and over the wire.]{Container-to-container packet paths using BPF redirects and \texttt{netkit} device pairs, on a single host (\pathid{2}) and over the wire (\pathid{3} and \pathid{1}).}
  \label{fig:netkit-packet-paths}
\end{figure}
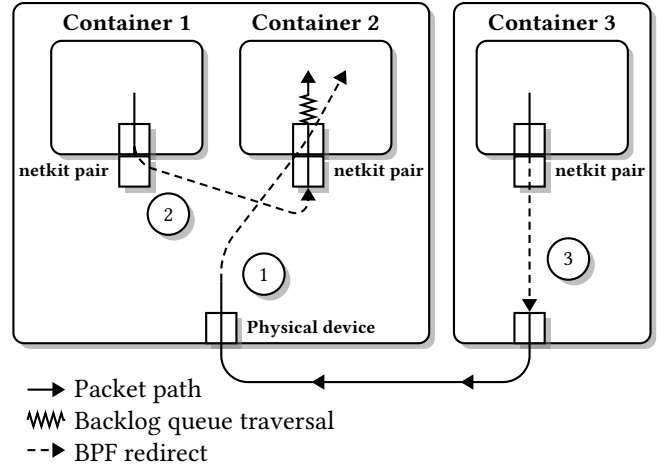

The design of netkit focuses on eliminating redundant backlog queue traversals during network namespace transitions.
We start by addressing the simplest case: transitions from the physical device to the container namespace.
Next, we introduce the new netkit device pairs required to remove backlog queue traversals when exiting containers.
Finally, we detail the remaining traversals that cannot be eliminated on ingress into the host and container, and explain why.
Figure \ref{fig:netkit-packet-paths} shows the different packet paths when using netkit.

\subsection{Cross Network Namespace Redirection}

\label{sec:bpf_redirect_peer}

We begin by addressing path \pathid{1} from Figure \ref{fig:vanilla-packet-paths}, which routes packets from the physical device to the container's upper networking stack.
This path goes through the per-CPU backlog when packets are received on the container's veth device, before being processed by the container's own upper networking stack.

The core intuition behind netkit's first optimization is that the receive-side code path of the physical device is functionally identical to the code path executed after the backlog queue of the veth device inside the container, albeit in a different network namespace.
In both cases, the Linux upper networking stack logic is executed, with the set of tc-bpf programs, netfilter rules, IP routes, etc. belonging to either the host or the container network namespace.
This redundancy presents an opportunity for optimization.

Our design avoids the backlog queue traversal in the container network namespace by redirecting packets from the upper stack of the host straight to the upper stack of the container.
This allows us to skip both the transmit code path of the veth device on the host side and the backlog queue traversal on the container side.
This redirection is illustrated as path \pathid{1} in Figure \ref{fig:netkit-packet-paths}.

This approach only works if the Linux network stack is not required to enforce security policies, maintain Quality of Service (QoS), or implement Network Address Translation (NAT).
These features are typically handled by either netfilter or BPF.
Therefore, netkit assumes that all such functionality is implemented in BPF before the redirection to the container's networking device.
They can be implemented at the tc-bpf or the XDP hooks.
In practice, this requires a container datapath that does not rely on netfilter in the host, as any rule there would be bypassed by the redirection.
This has become a reasonable assumption as BPF-based datapaths gained adoption.
The redirection also skips the host device, so any BPF program attached to it must move to the container's device or to the physical device.

This redirect is enabled by a new BPF redirect helper, \texttt{bpf\_redirect\_peer}, that switches the network namespace for the redirected packet.
This helper is called from the tc-bpf hook on the host, switches the packet to the network namespace of the target container, and recirculates it through the upper stack logic.

It does not exist at XDP because XDP operates prior to Generic Receive Offload (GRO) and thus processes individual, non-aggregated packets.
Implementing it there would require several redirects, whereas our design leverages GRO aggregation (including BIG TCP aggregations beyond 64KB), reducing the overhead to a single, consolidated packet.

\subsection{The netkit Device Pair}

Eliminating the backlog queue traversal when leaving the container's network namespace is trickier.
This backlog queue is illustrated on paths \pathid{2} and \pathid{3} of Figure \ref{fig:vanilla-packet-paths}.
We first focus on path \pathid{3}, specifically from the container to the physical device.

On this packet path, the backlog queue resides on the receive side of the host's veth device.
To avoid this queue, we would need to jump directly from the container's veth device to the physical device.
However, this approach requires installing a BPF program within the container's network namespace.
Such a program may not only conflict with the application's own BPF programs, but it could also be unloaded by the application.
Depending on how policies are enforced, unloading a program can either bypass security measures or disrupt connectivity.
In addition, a BPF program running in the network namespace of the container would have no access to information in the host namespace, such as the routing table.

To address these challenges, we introduce \texttt{netkit} devices, a new type of paired devices that connect containers to the host.
netkit devices treat BPF as a first-class citizen, enabling programs to be attached directly to the device without requiring to setup tc qdiscs and filters.
The BPF programs attached to a netkit device are executed as part of the driver's logic on the transmit path, immediately after the network namespace switch.
This design allows a BPF program attached to the container's netkit device to run in the context of the host's network namespace, enabling it to resolve IP routes from the host and redirect packets to physical devices.
netkit devices operate in layer 3 mode by default, which avoids the neighbor resolution that veth requires, and can be configured in layer 2 mode for containers that expect Ethernet semantics.

netkit devices deliberately do not support XDP: the performance benefits of XDP on virtual devices remain unclear, and supporting it adds significant code complexity, as seen in the veth driver \cite{veth-xdp}.
It also conflicts with our goal of running BPF programs on GRO-aggregated packets rather than on individual ones, and since container applications listen on regular IPv4/IPv6 sockets, packets seen by XDP must traverse the upper stack afterwards regardless.

Unlike veth devices, the two devices in a netkit pair are not interchangeable: one is designated as the host (or primary) device, while the other is the peer device.
The latter is intended to reside in the container's network namespace.
This distinction ensures that applications inside the container cannot attach or detach BPF programs on the peer device.
Instead, only the control plane, Cilium in our case, can manage them through the primary device, even though the peer programs execute on the transmit path of the peer netkit device. This allows us to run BPF programs on the transmit path of the container's device, before the backlog queue, while preventing containerized applications from tampering with them.
When no program is attached, both devices fall back to a default policy which can be set to drop, so that no traffic leaks before the control plane attaches its programs.

Thanks to this new netkit device, we can now use the standard BPF redirect helper to redirect packets from the container's netkit device straight to the physical device on the host, as illustrated on path \pathid{3} of Figure \ref{fig:netkit-packet-paths}.
Note we don't need our new helper here because netkit devices execute their BPF programs after the namespace switch.
Since no backlog queue is involved, the packet remains in the context of the sending process, preserving scheduler accounting and avoiding the ksoftirqd handoff.

Similarly, we avoid the backlog queue traversal on path \pathid{2} of Figure \ref{fig:vanilla-packet-paths} by redirecting from the host-side netkit device of the source container to the host-side netkit device of the destination container.
This redirection is illustrated on path \pathid{2} of Figure \ref{fig:netkit-packet-paths}.
We explain in Section \ref{sec:why-redirect-for-tx-to-rx} why we can't redirect inside the destination container.

\subsection{Required Backlog Queue}

As is apparent in Figure \ref{fig:netkit-packet-paths}, even with netkit devices, one backlog queue traversal remains.
This last traversal happens on the receive path of the destination container when both containers reside on the same host.
Unlike the other traversals discussed earlier, it cannot be eliminated without substantial refactoring of the Linux networking stack.

\label{sec:why-redirect-for-tx-to-rx}

For packets originating from another network namespace on the same host, at least one backlog queue traversal is required to transition from the transmit code path to the receive code path.
In our design, we keep the backlog queue traversal at the destination by redirecting traffic from the container's netkit device.
Alternatively, we could have kept the backlog queue traversal at the source by redirecting traffic after the host's backlog queue, using the tc-bpf hook of the host's netkit device and the helper introduced in Section \ref{sec:bpf_redirect_peer}.
We chose the former approach because it allows us to have a consistent attach point, netkit, instead of mixing netkit and tc-bpf hooks.
The only way to avoid it entirely would be to bypass the networking stack via sockmap, which again assumes applications do not require the full stack.

\section{Implementation \& Evaluation}

Through our evaluation, we aim to answer three key questions:
\begin{enumerate}
  \item How much effort is required to implement netkit and integrate it with existing container networking software?
  \item What performance gains does netkit achieve?
  \item How close are we to fully eliminating the overhead of network namespace transitions?
\end{enumerate}

\subsection{Implementation}

Our new BPF redirect helper required only 78 new lines of code, as we could reuse most of the logic from the existing \texttt{bpf\_redirect} helper.
The new netkit device, however, introduced 1,048 lines of code, though a significant portion of this is boilerplate code for BPF links, netlink, and kernel module creation.
The netkit device supports attaching multiple BPF programs via the \texttt{bpf\_mprog} API.
It supports the same program type as tc-bpf and can therefore use the same BPF helpers and kfuncs.

The new \texttt{bpf\_redirect\_peer} helper was upstreamed in Linux v5.10 \cite{redirectpeer-patch} and the netkit devices in v6.7 \cite{netkit-patch}.

Adding support in Cilium \cite{cilium} for our new helper, which by itself eliminates one backlog queue traversal, required 49 line changes\footnote{Line changes include added \textit{and} removed lines, but not comment additions.}, including logic to detect kernel support and fallback to the old redirection mechanism.
Support for the new device pairs is more involved, as it requires device creation and a shift from tc-bpf to the device-native attachment mechanism of netkit.
Implementing this in Cilium required 463 line changes.

Overall, we find that integrating netkit involves relatively straightforward changes in both the kernel and container networking software.
While using netkit devices demands more effort, it does not necessitate major architectural changes, as it aligns with the existing model of one device on each side of the namespace boundary.

\subsection{Network Performance}

To measure the performance impact of our changes, we deploy a two-node Kubernetes cluster with Cilium v1.19.5 handling connectivity.
The Kubernetes nodes run on the same hardware as in Section \ref{sec:background}, including the 3.1GHz CPU, with Linux v6.8.
To compile the BPF programs, LLVM v19.1.7 is used with the eBPF instruction set v3 \cite{ebpf-isa-extensions}.
eBPF JIT compilation is always enabled.

Cilium runs in native routing mode with BPF masquerading, the bandwidth manager, and kube-proxy replacement enabled.
kube-proxy and its netfilter rules are removed from all nodes.
As per the Cilium tuning guide \cite{cilium-tuning}, we disable Hubble and let Cilium bypass the netfilter connection tracking for Pod connections.
In this configuration, Cilium already bypasses netfilter in the host and relies on BPF to implement policy enforcement, QoS, or load balancing.
Our evaluations focus on the backlog queue traversals that netkit eliminates\footnote{When comparing to the host-to-host setup, netkit bypasses some additional connection tracking in Linux, as Cilium only bypasses connection tracking for Pod connections. We discuss the impact of this difference in Section \ref{sec:cpu_overhead}.}.

For the Host networking scenario, netperf runs in the host network namespace, communicating over the loopback device when on the same node.
For the other two scenarios, netperf runs in Kubernetes Pods.
When running the \texttt{TCP\_CRR} test, we increased the maximum size of the netfilter conntrack table to 300k entries, to ensure throughput isn't limited by the table size.

\begin{figure}
  \centering
  \includegraphics[width=\linewidth]{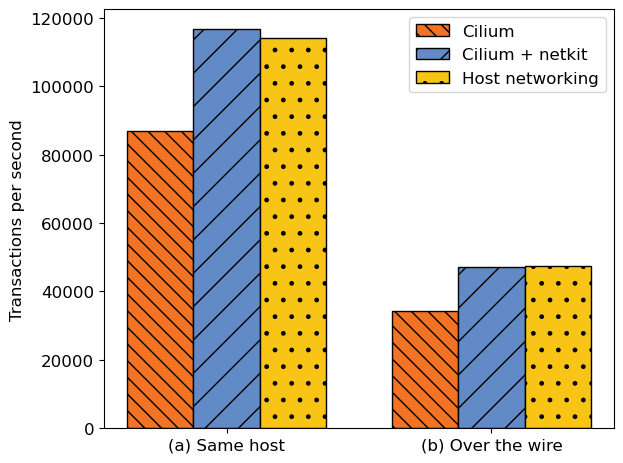}
  \Description{Transactions per second (\texttt{TCP\_RR} benchmark) between hosts, between containers connected with veth, and between containers connected with netkit.}
  \caption{Transactions per second (\texttt{TCP\_RR} benchmark) between hosts, between containers connected with veth, and between containers connected with netkit.}
  \label{fig:tcp_rr}
\end{figure}

We first look at the \texttt{TCP\_RR} results from Figure \ref{fig:tcp_rr}, which measure the request-response rate for small TCP packets.
This netperf test is an indirect measurement of the latency impact.
Throughput increases by 34\% on the same node and 37\% over the wire when relying on netkit for packet delivery in Cilium.
With netkit's improvement, Cilium achieves the same performance as two processes communicating without any container isolation.

\label{sec:netkit-host-diff}

A small difference remains between Cilium+netkit and the baseline, sometimes leading to Cilium+netkit performing better.
Comparing flamegraphs reveals two differences.
On one hand, Cilium+netkit can perform slightly worse because it takes longer to execute the BPF programs for its containers.
On the other hand, it can sometimes perform slightly better because it skips the remaining netfilter logic in the host.
Although visible in flamegraphs, these differences remain within our error margin.

\begin{figure}
  \centering
  \includegraphics[width=\linewidth]{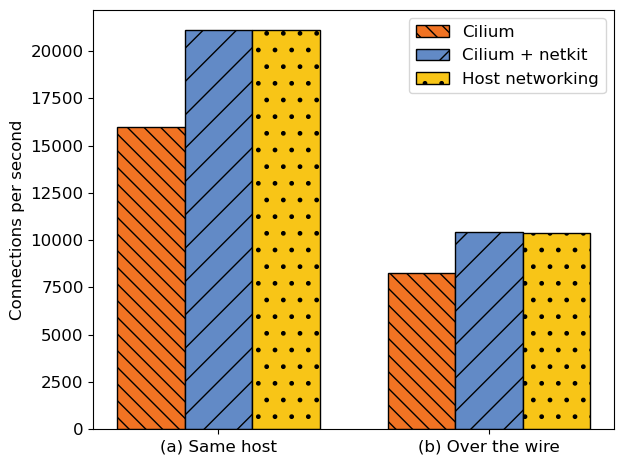}
  \Description{Connections per second (\texttt{TCP\_CRR} benchmark) between hosts, between containers connected with veth, and between containers connected with netkit.}
  \caption{Connections per second (\texttt{TCP\_CRR} benchmark) between hosts, between containers connected with veth, and between containers connected with netkit.}
  \label{fig:tcp_crr}
\end{figure}

We next look at the \texttt{TCP\_CRR} results from Figure \ref{fig:tcp_crr}, which measure the rate at which short connections can be established.
\texttt{TCP\_CRR} represents a more realistic workload, in particular when considering applications with many short-lived connections, such as IAM applications.
As shown in Figure \ref{fig:tcp_crr}, netkit has a similar impact as previously, increasing the connection rate by 32\% for containers on the same node and 26\% across the wire.

Not only does netkit significantly improve performance, but it also eliminates the overhead from network namespace transitions.
Over the wire, these results confirm that the extra backlog queue traversals account for most of the overhead.

\subsection{CPU Overhead}

\label{sec:cpu_overhead}

\begin{figure}
  \centering
  \includegraphics[width=\linewidth]{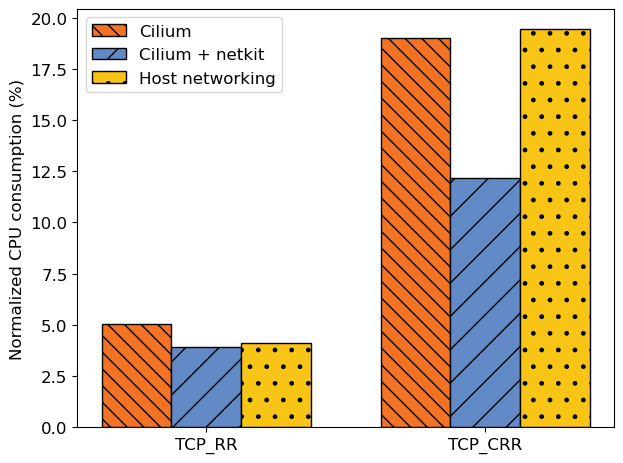}
  \Description{Normalized CPU consumption at the receiver for the \texttt{TCP\_RR} and \texttt{TCP\_CRR} tests over the wire.}
  \caption{Normalized CPU consumption at the receiver for the \texttt{TCP\_RR} and \texttt{TCP\_CRR} tests over the wire.}
  \label{fig:cpu}
\end{figure}

To complete our network performance numbers, we measure the CPU consumption, normalized to the same throughput, at the receiver for the \texttt{TCP\_RR} and \texttt{TCP\_CRR} tests over the wire.
Two points stand out from the results shown in Figure \ref{fig:cpu}.
First, the \texttt{TCP\_CRR} test stresses the networking stack a lot more than \texttt{TCP\_RR}.
This is expected as it involves many short TCP connections and therefore strains the connection tracking components in particular.

Second, the CPU consumption decreases when using netkit, even compared to the Host networking case.
This decrease is related to the remaining difference between Cilium+netkit and the Host networking discussed in Section \ref{sec:netkit-host-diff}.
When using netkit, we skip the connection tracking of the Linux kernel on the host.
Since this represents most of the overhead for the \texttt{TCP\_CRR} test, it significantly helps reduce CPU consumption.

\section{Related Work}

There is a large body of work around both container networking performance \cite{container-performance, spright, hybridmesh, canalmesh, sgiov} and BPF-driven performance improvements \cite{xdp, ovs-afxdp, dwivedi25}.
Several papers \cite{spright, canalmesh, hybridmesh} sit at the intersection of both, using diverse eBPF hooks to accelerate communications in the context of service meshes and their sidecar proxies.
In particular, their use of socket-level redirections most closely resembles our work.
Socket-level redirections however are not a fitting solution for container applications that expect a full network stack, and as noticed in \cite{canalmesh}, it suffers from its own performance shortcomings.

netkit shares similarities with \texttt{ipvlan} \cite{ipvlan} in Linux.
In ipvlan-based deployments, each container is assigned its own ipvlan slave device, all of which are connected to a single master physical device on the host.
ipvlan also eliminates the backlog queue traversal on ingress into the containers and bypasses part of the host networking stack.
It however comes with a number of drawbacks.
First, the host networking stack is not bypassed on egress, which can lead to path asymmetry issues\footnote{This problem is addressed by the L3S mode, but with reduced performance.}.
Second, BPF programs attached to the ipvlan slave devices can be removed by container applications, and they have no access to information from the host network namespace.
Third, an ipvlan slave is bound to a single master at creation time, so multiple uplinks require one slave each or an aggregation device such as a bond.
Fourth, ipvlan maintains an internal FIB, which is less flexible for BPF programmability than reusing the kernel FIB via \texttt{bpf\_fib\_lookup}.
All of these issues have led the Cilium community to remove support for ipvlan devices \cite{cilium-ipvlan}.

Outside of container networking, several papers discuss the design and use of new high-performance packet delivery mechanisms for Linux.
XDP \cite{xdp} enables early packet processing directly in Ethernet drivers, while AF\_XDP \cite{ovs-afxdp} extends this capability by delivering packets to userspace via a new socket type.
Although XDP can be used in container networks, AF\_XDP is less practical, as it requires applications to adopt the new socket interface.
It also has no native container integration: a container can only bind such a socket to its veth device.
That device lacks zero-copy support and must convert already allocated skbs back into XDP buffers, losing the bypass that makes AF\_XDP fast on physical devices.

\section{Conclusion}

netkit demonstrates that the overhead of network namespace transitions in containerized environments can be entirely eliminated by specializing the Linux networking stack with eBPF.
By avoiding redundant backlog queue traversals and introducing a new device pair that treats BPF as a first-class citizen, netkit achieves up to 37\% throughput improvements and matches the performance of process-to-process communication, all while requiring minimal changes to existing container networking software.

Since its introduction to the Linux kernel, netkit has been deployed in production at scale by Meta \cite{meta-user} and ByteDance \cite{bytedance-user} to improve container networking performance.

\section{Acknowledgments}

We thank Kahina Lazri and the anonymous reviewers for their valuable feedback.

\bibliography{references}

\end{document}